# Changing Geographies of Violence in North and West Africa

Steven M. Radil[1,2], steven.radil@ufl.edu
Olivier J. Walther[1], owalther@ufl.edu
Mohammed Sanda[1], msanda1@ufl.edu

[1] Department of Geography, University of Florida
[2] Corvinus Institute for Advanced Studies (CIAS), Corvinus University of Budapest

10/9/2026

**Abstract**
This report analyses spatial trends in armed conflicts in North and West Africa. It shows, first, that all types of violence saw a significant increase in 2025. However, these violent incidents are distributed very unevenly across the region: while conflict is at its lowest in North Africa, the vast majority of violent incidents and victims are concentrated in five West African countries: Nigeria, Cameroon, Mali, Burkina Faso, and Niger. Coastal countries are experiencing a very mixed picture: jihadism is not advancing as a front, from the Sahel toward the Gulf of Guinea, and most of the violence stems from local disputes, ethnic conflicts, and banditry. Second, the spatial indicator developed by the OECD/SWAC to track the evolution of this violence confirms that the conflict landscape is expanding at its margins and intensifying at its centre. Even more concerning is that peripheral areas are increasingly being drawn directly into well-established conflict zones, rather than going through a gradual, low-intensity transition phase. The geography of expansion is shifting from a diffuse, low-intensity spread toward peripheral areas to a rapid outward expansion from dense, well-established conflict hotspots. The combination of increasing intensity and stable concentration indicates that violence is becoming more frequent in familiar locations without fundamentally altering its spatial distribution within each cluster. The local indicator used in the report suggests that violence continues to spread to new locations at the margins of established conflict zones. At the heart of these zones, tensions are intensifying relative to their historical baseline levels, and a growing number of regions have experienced clashes for a decade or more. Our projections indicate that the intensity of violence is expected to increase in 2026, reaching more than 15,000 incidents and more than 32,500 deaths. Much of the increase is explained by the growing number of conflicts in Nigeria. In the Sahel, fatalities should increase in Mali and decrease in Burkina Faso.

**Acknowledgments**
This work was funded by the OECD Sahel and West Africa Club.

## Violence in North and West Africa

Political violence in North and West Africa[1] has continued to evolve along divergent trajectories (Figure 1). North of the Sahara, violence remains at historically low levels compared with the peak years of the Libyan civil wars (2011, 2014-2020), though a possible reversal is now detected. In 2025, 413 violent events resulting in 360 fatalities were recorded across Algeria, Libya, Morocco, and Tunisia, a 63% increase in events over 2024. A decade earlier, the same region recorded 1,288 events and 3,365 deaths. The 2025 increase was driven by Libya, where events rose from 203 to 352, the highest annual total since the permanent ceasefire of 2020. Much of the increase is due to clashes between militias around Tripoli and conflict between the Libyan National Army (LNA) and Tebu armed groups in Al Kufrah region. Even so, violence in North Africa accounts for less than 4% of the events observed across the broader region and remains qualitatively different from the large-scale armed conflicts further south.

**Figure 1.** Violent events and fatalities in North and West Africa, 1997-2025

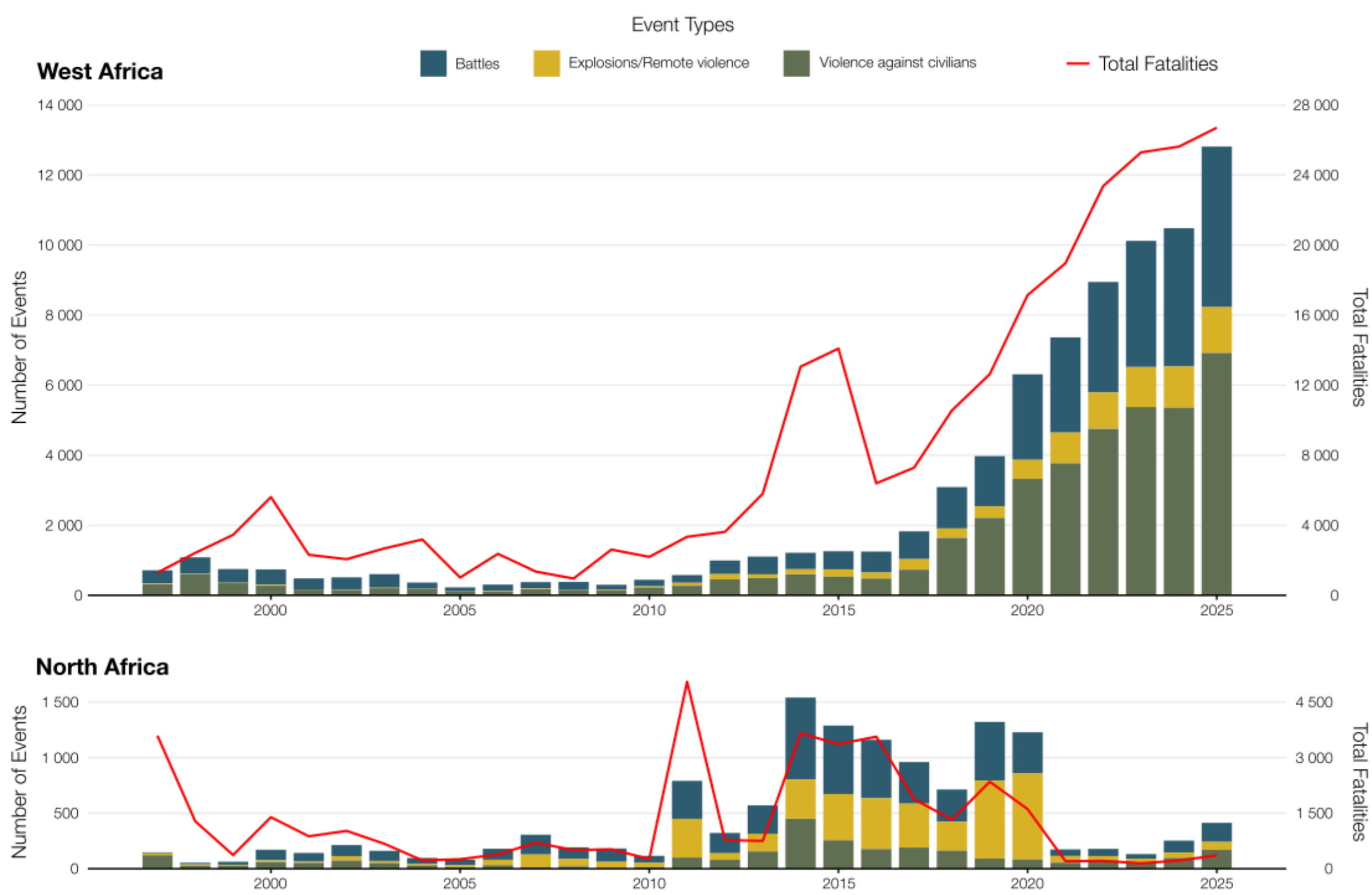

Source: Authors, based on ACLED (2026).

[1] The report study region is defined as the 15 countries typically identified as part of West Africa: Benin, Burkina Faso, Cabo Verde, Côte d'Ivoire, Ghana, Guinea, Guinea-Bissau, Liberia, Mauritania, Niger, Nigeria, Senegal, Sierra Leone, The Gambia, and Togo. A further set of seven surrounding countries that share international land boundaries with the above list of countries are included given the propensity for conflict to cross boundaries and affect borderlands: Algeria, Cameroon, Chad, Libya, Morocco, Tunisia, and Western Sahara. Results for Western Sahara are combined with Morocco given the latter's de facto control over most of Western Sahara's territory.

South of the Sahara, 2025 was the most violent year recorded since data collection began in 1997. West Africa experienced 12,811 violent events and 26,699 fatalities, increases of 22% and 4% respectively over 2024. More than 218,000 people have now been killed in the region since the beginning of the Boko Haram insurgency in 2009. All three categories of violence tracked by the Armed Conflict Location and Event Data (ACLED) project increased in 2025. Violence against civilians remained the most common event type, with 6,918 incidents and 8,400 fatalities, compared with 610 incidents in 2014. Battles between armed groups, which include government forces and non-state actors, produced 15,341 fatalities in 2025, nearly five times the toll recorded in 2017, the year Jama'at Nusrat al-Islam wal-Muslimin or Group for the Support of Islam and Muslims (JNIM) was formed. Explosions and remote violence, a category that includes Improvised Explosive Devices (IED) attacks, airstrikes, and drone strikes, accounted for 1,309 events and 2,958 fatalities.

The gap between the growth in events and in fatalities reflects a declining per-event lethality across West Africa. In 2024, each violent event produced an average of 2.4 fatalities. In 2025, that figure fell to 2.1, a 15% decline. This pattern is consistent with conflict spreading into new locations where violence can be frequent but less intense, while the core conflict locations remain extraordinarily deadly. The Central Sahelian conflict zone (comprised of Mali, Burkina Faso, and the western departments of Niger) for example, accounts for 26% of regional events but 44% of fatalities.

*Two hotspots of violence*

Violence in the region is concentrated in a small number of countries. Just four West African countries (Nigeria, Mali, Burkina Faso, and Niger) plus Cameroon accounted for 92% of violent events and 95% of fatalities during 2025 (Table 1). The remaining countries in the study area experienced comparatively low levels of violence, though a few are on trajectories of concern.

The Central Sahelian conflict zone remains the largest hotspot of violence in the region. A staggering 10,467 people were killed in these areas across 3,243 incidents during 2025, representing 39% of all fatalities and 25% of all events recorded in North and West Africa. The three countries have experienced different trajectories in 2025, however. In Burkina Faso, events increased by only 4% (from 1,310 to 1,362) while fatalities dropped by 30% (from 7,518 to 5,263). Mali followed a broadly similar pattern, with events rising 8% and fatalities declining slightly. In western Niger's Tillabéri region, 297 events produced 1,276 fatalities during 2025.

The sustained violence across Burkina Faso in particular has continued to engulf much of the national territory, extending well beyond the Sahel, Est, and Centre-Nord regions that were the primary zones of conflict in 2021-22 (OECD/SWAC, 2023). The expansion has reached the regions bordering Côte d'Ivoire, Ghana, and Togo in the south. Rural areas and small urban centres bear the brunt of the violence, while the capital Ouagadougou and other large cities have been largely spared, a pattern of ruralisation of conflict that contributes to the growing isolation of major cities and the fragmentation of national territory (Radil et al., 2023).

**Table 1.** Violent events and fatalities across North and West Africa, 2025

| | Violent events | | Fatalities | | Population | | Events (%) vs population (%) | Fatalities (%) vs population (%) |
|---|---|---|---|---|---|---|---|---|
| | Number | % | Number | % | Number | % | Ratio | Ratio |
| Nigeria | 5 244 | 39.7 | 12 630 | 46.7 | 237 527 782 | 38.2 | 1.0 | 1.2 |
| Cameroon | 3 401 | 25.7 | 1 857 | 6.9 | 29 879 337 | 4.8 | 5.4 | 1.4 |
| Mali | 1 584 | 12.0 | 3 928 | 14.5 | 25 198 822 | 4.0 | 3.0 | 3.6 |
| Burkina Faso | 1 362 | 10.3 | 5 263 | 19.5 | 24 074 580 | 3.9 | 2.6 | 5.0 |
| Niger | 530 | 4.0 | 1 905 | 7.0 | 27 917 831 | 4.5 | 0.9 | 1.6 |
| Libya | 352 | 2.7 | 276 | 1.0 | 7 458 555 | 1.2 | 2.3 | 0.8 |
| Ghana | 248 | 1.9 | 325 | 1.2 | 35 064 273 | 5.6 | 0.3 | 0.2 |
| Benin | 153 | 1.2 | 402 | 1.5 | 14 814 460 | 2.4 | 0.5 | 0.6 |
| Chad | 124 | 0.9 | 247 | 0.9 | 21 003 705 | 3.4 | 0.3 | 0.3 |
| Togo | 85 | 0.6 | 91 | 0.3 | 9 721 608 | 1.6 | 0.4 | 0.2 |
| Others | 141 | 1.1 | 135 | 0.5 | 189 895 569 | 30.5 | 0.0 | 0.0 |
| **Total** | **13 224** | **100.0** | **27 059** | **100.0** | **622 556 522** | **100.0** | **1.0** | **1.0** |

Sources: Authors, based on ACLED (2026) and United Nations (2025) data.
Notes: Data are for calendar year 2025. In countries with a ratio higher than 1, violent events or resulting fatalities are higher than the share of the regional population would suggest. Population data are mid-year 2025 estimates from the United Nations World Population Prospects, 2024 revision.

The second major hotspot is Nigeria, which recorded 5,244 events and 12,630 fatalities during 2025, representing 39% of the regional events and 47% of fatalities. Nigeria's internal geography of violence has also continued to shift. The Northwestern states of Kaduna, Katsina, and Zamfara now account for 32% of the country's events and 30% of its fatalities, exceeding or matching Borno state, the core of the Boko Haram/Islamic State West Africa Province (ISWAP) insurgency, which recorded 13% of events and 31% of fatalities. The Northwestern zone is affected by a convergence of organised banditry, communal violence, and an increasingly visible jihadist presence that complicates the distinction between criminal and politically motivated violence (Madueke et al., 2024; Ojewale, 2024). Borno, the historical centre of the Boko Haram and ISWAP insurgency, remains the single most lethal state, with 3,856 fatalities in 2025. Other major concentrations of violence include the Middle Belt (Benue with 876 fatalities, Plateau with 557), Niger State (607 fatalities), and Sokoto (570 fatalities). Nigeria's conflicts have coalesced into a single vast zone of violence that stretches from the Lake Chad Basin to the Niger Delta, encompassing the greater part of the country's territory (Map 1).

Map 1. Violent events in North and West Africa, 2025

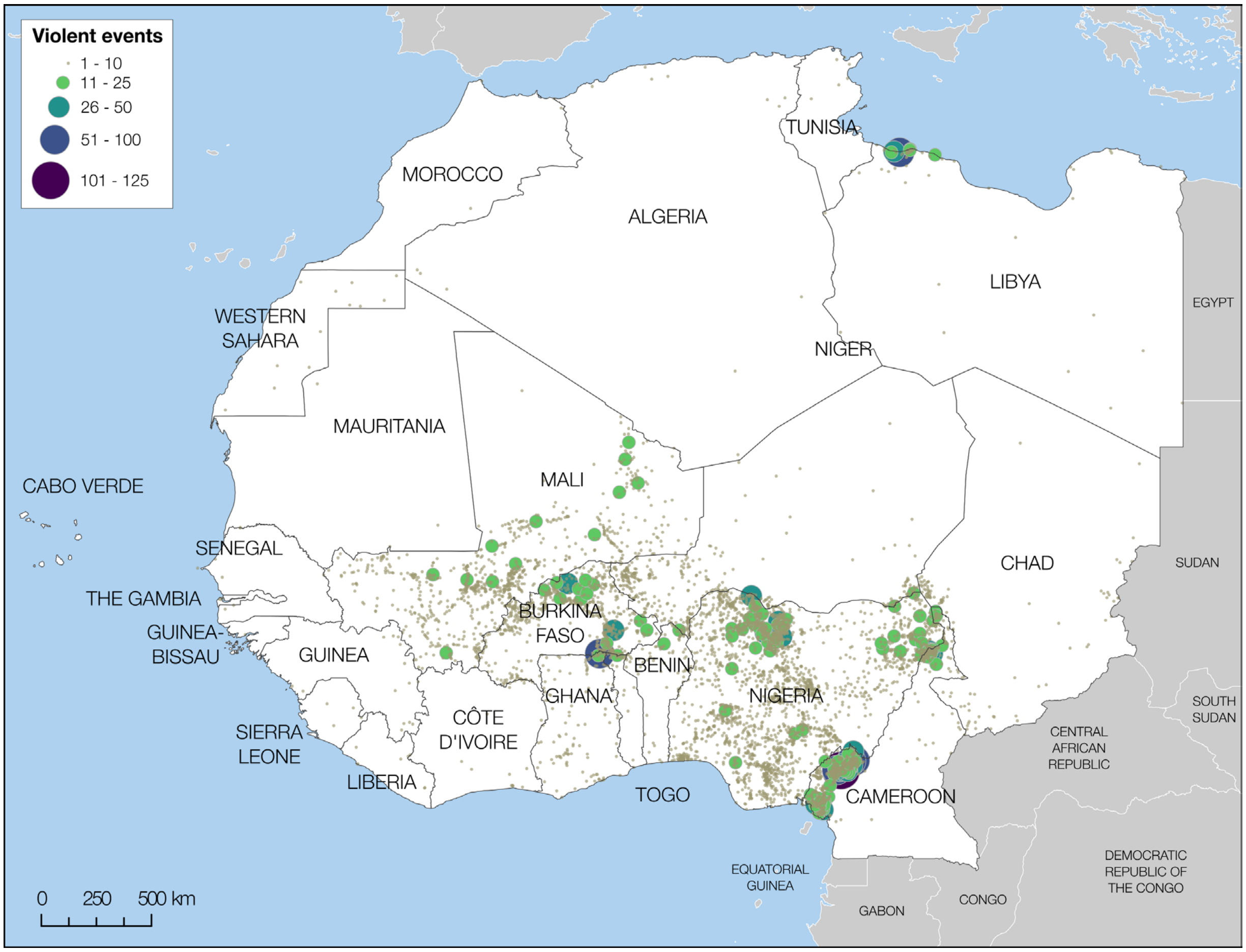

Source: Authors, based on ACLED (2026).

The intensity of violence in each country correlates imperfectly with population size (Table 1). In countries where the ratio of event or fatality share to population share exceeds 1.0, violence is disproportionate to the country's demographic weight in the region. The most extreme examples are found in Cameroon, Burkina Faso, and Mali. Cameroon accounts for 26% of the region's events with less than 5% of its population, an events-to-population ratio of 5.4. Burkina Faso's disproportion is most visible on the fatality side: it accounts for 24% of all deaths with 4% of the regional population, a fatalities-to-population ratio of 5.0. Mali follows a similar pattern, with a fatalities-to-population ratio of 3.6. Nigeria is an exception in the opposite direction. As the most populous country in Africa, it accounts for 38% of the regional population and a similar share of events (39%), though a higher share of fatalities (47%). These ratios must be interpreted cautiously, however as the most recent Nigerian census was conducted in 2006, and the country's population figures carry substantial uncertainty. Further, ACLED's fatality counts should be interpreted as conservative minimum estimates; actual fatalities may be far higher (ACLED, 2026).

*Cameroon's continued crisis*

For the second consecutive year, Cameroon recorded the second highest number of violent events in the study region, with 3,401 incidents during 2025, a new annual record. The country's two major conflicts continue to drive this trajectory. The Anglophone Crisis, centred on the Nord-Ouest and Sud-Ouest regions, accounts for 76% of the country's violent events and 69% of its fatalities during 2025. The remainder is concentrated in the Extrême-Nord region, where the Boko Haram and ISWAP insurgency has long been active around the Lake Chad Basin.

Cameroon presents a distinctive conflict profile. Despite producing the second highest event count in the region, the number of fatalities (1,857) is substantially lower than in Burkina Faso (5,263) or Mali (3,928), both of which recorded fewer events. The Anglophone Crisis involves a high frequency of incidents at very low per-event lethality levels, a pattern consistent with the guerrilla tactics and reprisals that characterise this conflict. An additional feature is the close correspondence between the distribution of violence and the country's linguistic borders, a pattern rarely observed in other West African conflicts (Nikolovsky and Walther, 2025).

A concerning trend in Cameroon is the steady year-over-year increase in events, from 1,526 in 2022 to 3,401 in 2025. Yet fatalities have moved in the opposite direction, declining from 2,185 in 2024 to 1,857 in 2025. As with the broader West African pattern, each individual incident in Cameroon is becoming less lethal on average even as events proliferate overall.

*Jihadist expansion toward the coast*

The southward expansion of violence from the Sahel into the coastal countries of Benin, Côte d'Ivoire, Ghana, and Togo has continued at an uneven pace. These four countries recorded their highest combined annual totals of events and fatalities in 2025 since tracking began in the region in the late 1990s. Nearly 60% of all events recorded in these countries since 2014 have occurred since 2020, a reflection of how recently politically motivated violence has become a feature of the coastal zone.

The most significant development in 2025 was the sharp escalation in Ghana (see Figure 2). Unlike in Benin, where violence remains concentrated in the northern departments of Alibori and Atacora (75% of events, 81% of fatalities since 2014), Ghana's violence is somewhat more geographically dispersed. The Greater Accra region in the south was responsible for 15% of all events and fatalities. Much of these violent events are clashes between rival groups during traditional ceremonies or gang-related activities. The Upper East region along the border with Burkina Faso accounted for 29% of events and 21% of fatalities, much of it involving chieftaincy disputes, ethnic militia activity, and armed banditry rather than jihadist expansion. While JNIM has been reported as present in northern Ghana, as of mid-2024, no confirmed jihadist attack had occurred in Ghana. However, JNIM's cross-border activities have been linked to the Bawku interethnic conflict, raising concerns about the porous nature of the northern border (Salifu and Radil 2026).

**Figure 2.** Changes in events and fatalities in select coastal countries, 2024-2025

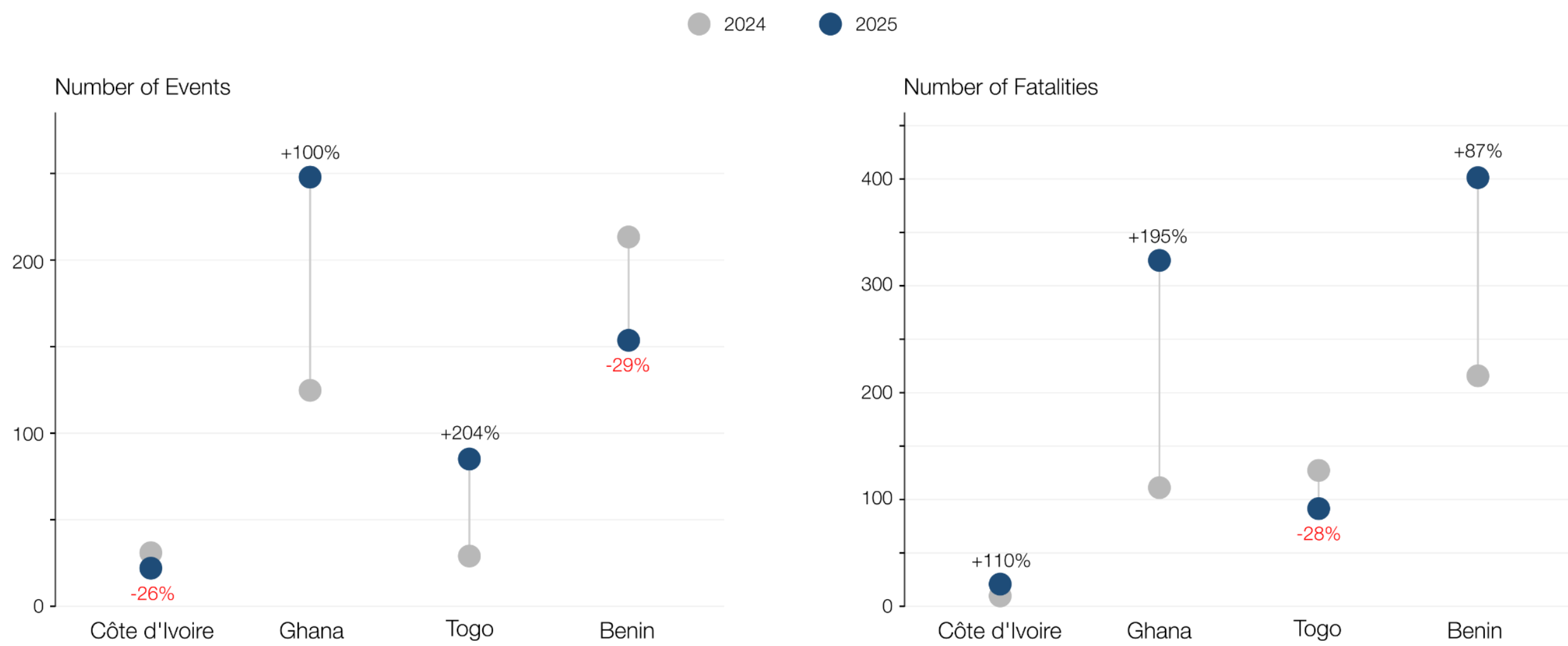

Source: Authors, based on ACLED (2026).

Benin presents a contrasting pattern. While the frequency of events declined in 2025, the incidents became significantly more deadly (Figure 2). Abductions, a signature feature of JNIM's activity in Benin's northern departments, declined from 58 in 2024 to 24 in 2025, though 88% of abductions remain concentrated in Alibori and Atacora. Northern Benin has emerged as the primary zone of armed confrontation between government forces and jihadist groups operating in the broader W-Arly-Pendjari (WAP) complex, a transnational natural park area spanning Burkina Faso, Benin, and Niger.

In Togo, event frequency surged from a small base, even as the resulting fatalities declined (Figure 2). Côte d'Ivoire continued to experience relatively few events, a continuation of the comparatively low violence levels that have followed the government's two-pronged approach of security operations and social investment in the northern regions (ICG, 2023b).

The analysis of conflict data over the past decade suggests an absence of a clear front progressing southward from the Sahel toward the Gulf of Guinea. Sporadic attacks continue to target security forces and civilians, especially in border zones, against the backdrop of a deepening jihadist presence in the northern areas of coastal countries. While the 2025 data shows alarming percentage increases in events in Ghana and Togo, recent research into the trajectories of violence also demonstrates that conditions in the conflict periphery are inherently unstable (Radil et al., 2026). Outbreaks in these newly affected zones frequently manifest as short-lived, ephemeral episodes rather than as entrenched conflict, meaning that rapid escalations in a single year can be followed by equally sharp contractions. This may indicate that jihadist groups are still in a building and recruitment phase in some coastal countries, while in others they view the coastal zone primarily as a logistics and transit corridor rather than a theatre of military operations.

## The Spatial Conflict Dynamics Indicator

The Spatial Conflict Dynamics indicator (SCDi) was developed by OECD/SWAC to monitor the changing geography of political violence in North and West Africa over time (OECD/SWAC, 2020; Walther et al., 2023a). The indicator has been updated annually as part of the West African Studies and West African Papers series (OECD/SWAC, 2020; 2021; 2023; 2025; Radil and Walther, 2024).

Where conventional conflict statistics report violence by country or administrative region, the base SCDi operates on a uniform grid that is independent of national or administrative boundaries. This design allows the indicator to capture cross-border dynamics and to compare conflict conditions across locations that differ widely in size, population, and political organisation. The SCDi divides the study area into 6,540 cells of 50 km by 50 km, covering a set of 21 countries in North and West Africa. Each cell is classified annually using disaggregated conflict data from ACLED. Three types of ACLED events are included in the analysis: battles between armed groups, explosions and remote violence (including IED attacks, airstrikes, and drone strikes), and violence against civilians. Other categories tracked by ACLED, such as protests and riots, are excluded because they do not consistently involve armed political violence.

The indicator measures two properties of violence in each cell: the intensity of conflict and the spatial concentration of conflict events relative to each other (Box 1). The combination of these two dimensions produces four conflict types (Figure 3), each with distinct implications for how armed groups operate, how populations are affected, and how security forces might respond.

**Figure 3:** How the SCDi uses event distribution and intensity to identify conflict categories

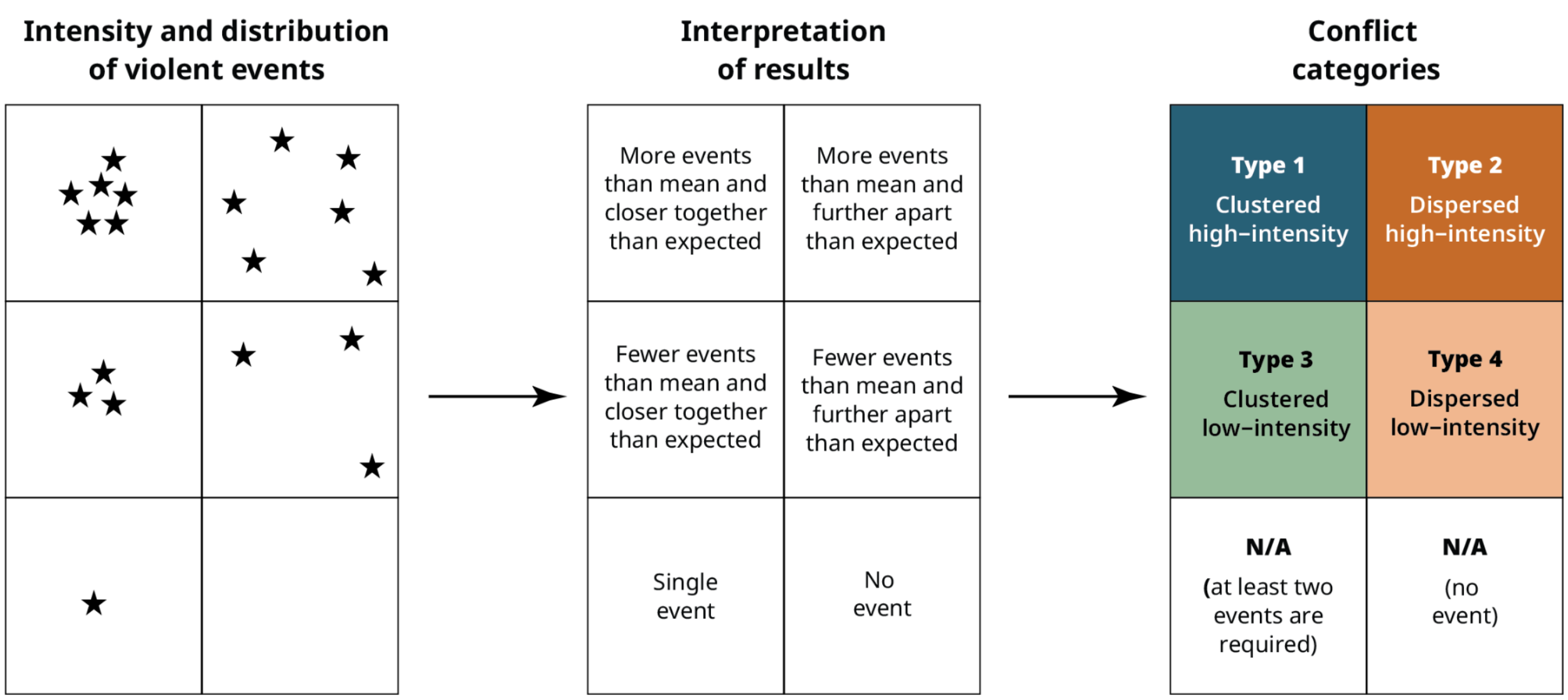

Source: OECD/SWAC (2020).

Clustered high-intensity cells (Type 1) are typical of the core zones of established armed conflicts, where fighting repeatedly occurs in and around the same locations. The Inner Niger

Delta in Mali, much of Burkina Faso, and the northeastern and northern states of Nigeria all contain extensive CH zones. These cells tend to record the highest numbers of fatalities.

Dispersed high-intensity cells (Type 2) are characteristic of areas where armed groups are actively expanding, attacking multiple targets across a wider area within the cell. The western bank of the Niger River in Niger, between Ayorou and Say, is one example. Cells classified as DH can signal a conflict that is becoming more geographically extensive.

Clustered low-intensity cells (Type 3) are typical of areas in transition, where conflict may be either building toward a more intense phase or dissipating. Episodes of violence in CL cells are often short-lived and followed by periods of relative calm. CL cells are commonly found at the margins of major conflict zones, in locations where armed groups are beginning to establish a presence, as in northern Benin, or where violence is winding down.

Dispersed low-intensity cells (Type 4) may indicate a lingering conflict that has not fully resolved, or isolated incidents of violence with no clear spatial pattern. Opponents in these areas may be highly mobile or unlikely to face protracted opposition in any single locality. Outside Nigeria, DL cells tend to appear in relative isolation from other conflict zones.

Cells that record no qualifying violent events in a given year, or that record only a single event (to be calculated, the conflict concentration metric requires at least two events), receive no SCDi classification and are not counted as conflict cells for that year. The SCDi time series runs from 1997, the first year of systematic ACLED coverage in the region, through December 2025.

Box 1. Measuring conflict intensity and concentration

Conflict intensity (CI) is defined as the total number of violent events in each cell for a given year, divided by the area of the cell (2,500 square km). The resulting metric equals zero if there are no events within a cell during a given year and has no upper limit. The SCDi classifies each cell as high or low intensity by comparing its CI score to an expected value called the "generational mean," which corresponds to the 20-year average CI between 1997 and 2016 for all cells that experienced conflict during that period. The generational mean for North and West Africa is 0.0017 events per square km per year, equivalent to approximately four events in a single year for a 50 by 50 km cell. A cell is classified as high intensity if four or more events occur in a given year, and as low intensity otherwise.

Conflict concentration (CC) measures the distribution of conflict locations relative to each other within a cell. An average nearest neighbour (ANN) ratio is calculated by dividing the observed average distance between events in a cell by the expected average distance if the same number of events were randomly distributed throughout the cell. ANN ratios smaller than 1.0 indicate clustering (events are closer together than random chance would predict), while ratios greater than 1.0 indicate dispersion. Cells with an ANN ratio below 1.0 are classified as clustered; those at or above 1.0 are classified as dispersed.

Source: Walther et al. (2023a).

*Interpreting the SCDi*

Several features of the SCDi's design are worth keeping in mind when reading the results that follow. First, the 50 km resolution means that the indicator captures regional patterns rather than individual incidents. A single cell covers an area roughly the size of a large rural department or a small province, and several events may fall within the same cell without being directly related to each other. The SCDi is designed to identify broad geographic shifts in conflict, not to track outcomes related to specific armed groups or local disputes.

Second, because the classification operates on annual data, the SCDi tracks medium-term trends rather than short-term fluctuations. A cell that shifts from no conflict to CL classification may be experiencing its first violence in a generation or may be seeing a return of violence after a period of calm. The base SCDi classification cannot distinguish between these two scenarios on its own. Supplementary metrics developed for recent editions of this analysis address this limitation by identifying cells that are newly entering conflict, tracking how many years of conflict a cell has accumulated, and comparing current conditions against each cell's own historical baseline (Radil and Walther, 2024). These local metrics are discussed later in this report.

The four SCDi types describe different spatial patterns of violence, each with distinct analytical implications, and should not be read as a severity ranking. CH cells point to entrenched, localised conflicts; DH cells may signal a conflict in geographic expansion. CL cells can represent either the early stages of a new conflict or the fading tail of an old one, while DL cells represent regions where conflict is lingering. Reading the SCDi effectively requires attention to the spatial relationships between cells, whether they form continuous zones, linear corridors, or isolated pockets, and to how individual cells change classification over successive years.

## The Ongoing Intensification and Spread of Violence

The SCDi shows that the geographic footprint of political violence in North and West Africa expanded substantially in 2025 after two years of very slow growth. A total of 805 cells were classified into one of the four SCDi conflict types, a new high in the 29-year time series and a 17% increase over 2024 (Figure 4). Whether this represents a resumption of the rapid expansion observed from 2016 to 2022, when the total nearly tripled, or a one-year spike driven by specific events, will only become apparent with subsequent updates.

**Figure 4.** Number of conflict zones by type, 1997–2025

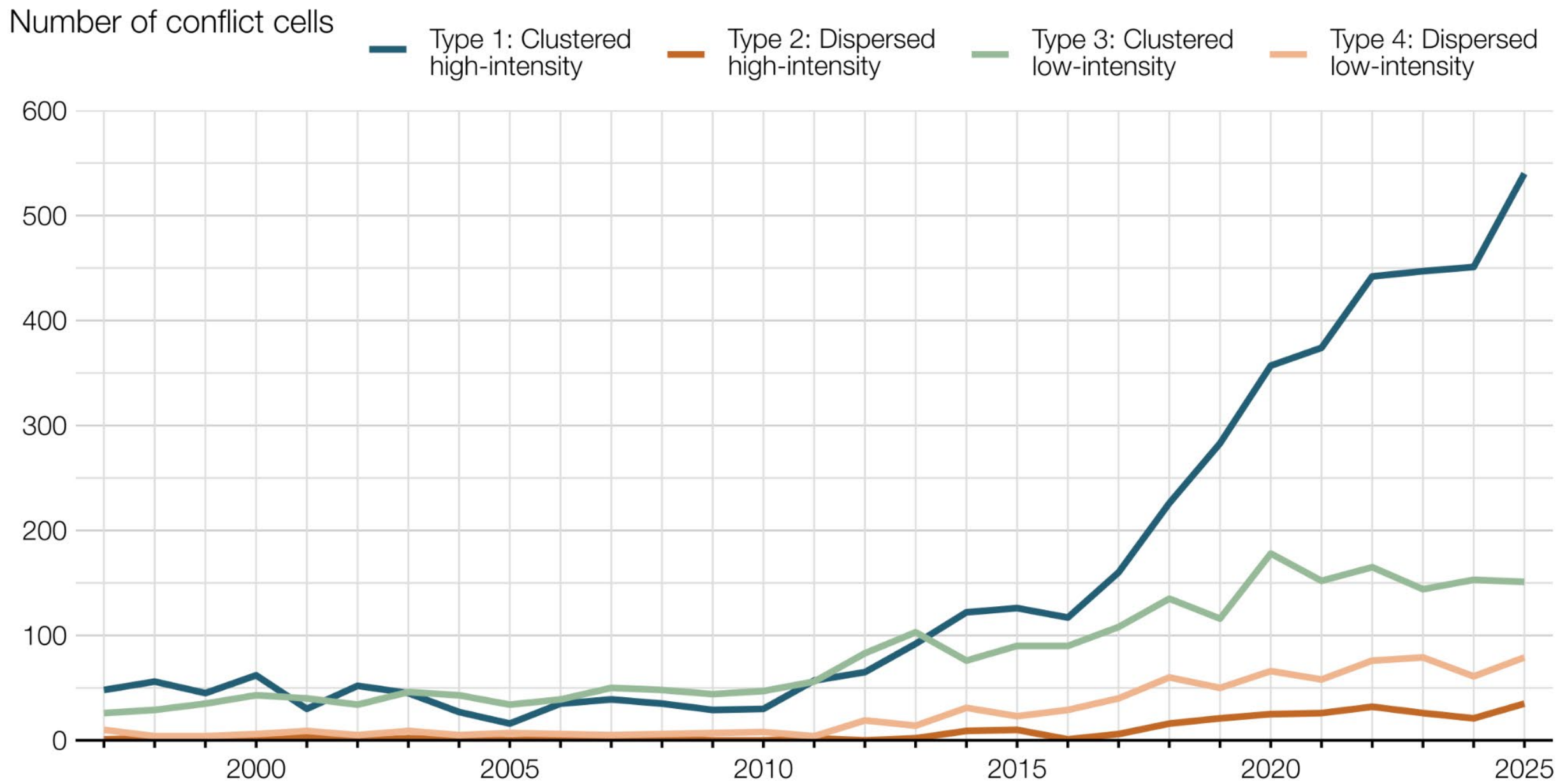

Source: Authors, based on ACLED (2026).

The composition of the 805 conflict cells in 2025 reflects an intensification of patterns that have been building for nearly a decade. Clustered high-intensity cells (CH) accounted for 67% of the total, continuing its annual increase since 2016 and reaching the highest share ever recorded (Figure 4). The growth of the CH category carries a clear analytical implication. The conflict map is increasingly dominated by cells where violence is both frequent and spatially concentrated, the signature of an entrenched armed conflict. These core zones are also growing while the transitional periphery expands.

The other three types of conflict regions have been more stable. Clustered low-intensity cells (CL) cells have held roughly steady at around 150 since 2023, occupying the edges of established conflict zones where violence may be either building or fading. Dispersed low-intensity cells (DL) cells returned to their 2023 levels after a decline in 2023. Finally, dispersed high-intensity (DH) cells, while few in number, increased notably and may indicate accelerating dynamics in specific locations.

*The declining share of low-intensity conflict*

As a consequence of the increase in clustered high intensity (CH) cells, the proportion of conflict cells classified as low-intensity (CL + DL) dropped below 30% for the first time in 2025, reaching 29% and extending a decline that has been nearly continuous since peaking above 70% in 2005 (Figure 5).

**Figure 5.** Conflict zones classified as clustered or low-intensity, 1997-2025

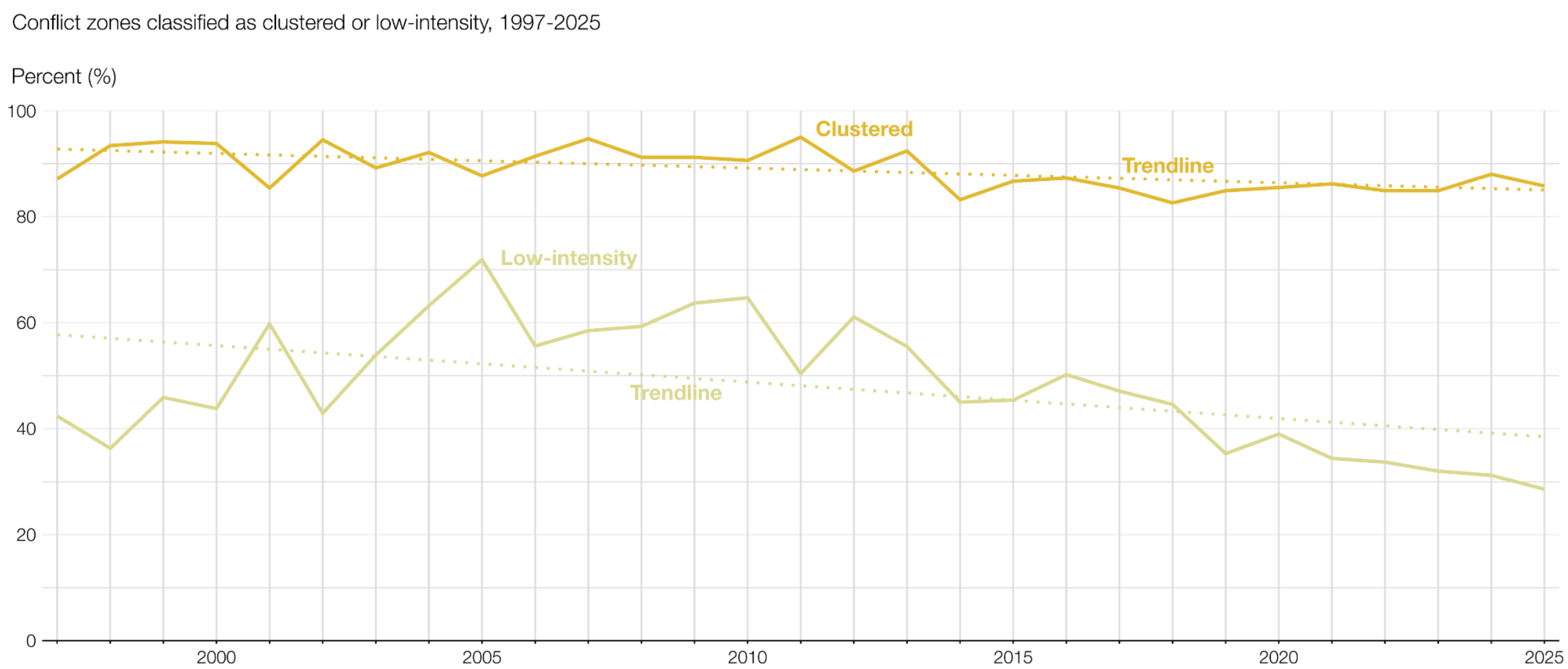

Source: Authors, based on ACLED (2026).

The significance of this trend lies in the relationship between the geographic spread of violence to new locations and local intensification. In earlier years, the expansion of violence to new locations pulled the low-intensity share upward, because cells entering conflict for the first time typically exhibited low event densities. The sustained decline observed in later years indicates that intensification within existing conflict zones has outpaced the rate of geographic expansion. Each year, a larger share of the conflict map consists of cells with recurring, high-frequency violence, while the proportion experiencing sporadic or transitional violence contracts.

The clustered share, by contrast, has remained largely stable. At 86% in 2025, it remains within the narrow band where it has fluctuated since 2014. Violence in the region continues to recur at the same locations within each cell, a pattern that has been essentially unchanged for over a decade.

**Map 2.** Spatial Conflict Dynamics indicator (SCDi) in North and West Africa, 2025

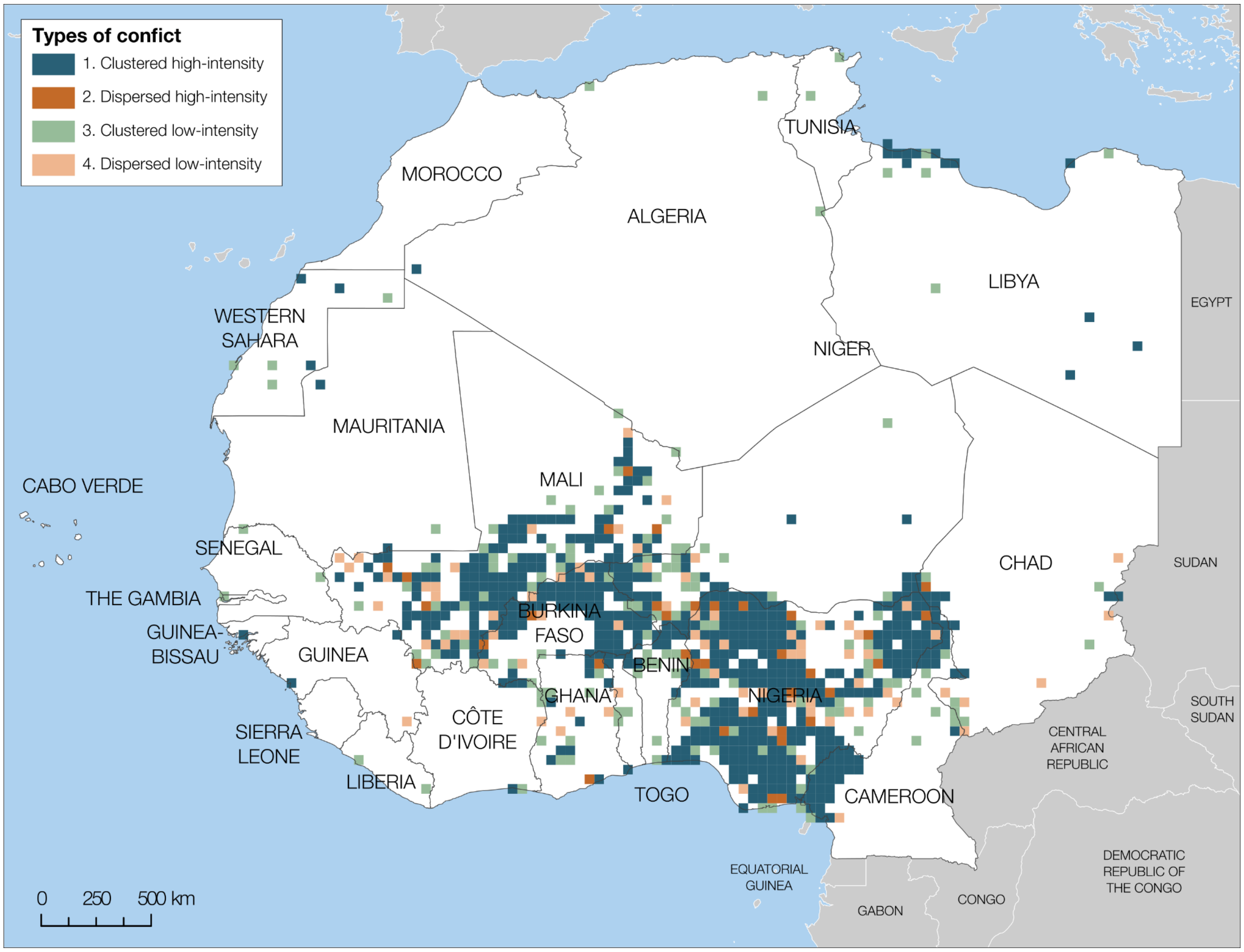

Source: Authors, based on ACLED (2026).

*How cells enter and move through conflict*

Of the 805 cells in conflict in 2025, 256 were not in conflict the previous year. Another 137 cells exited conflict over the same period, producing a net increase of 119 cells. The scale of turnover is large relative to the net change: roughly a third of the 2025 conflict map consists of cells whose status shifted in one direction or the other since 2024.

The composition of newly entering cells shows divergence from historical norms (Figure 6). Sequence analysis of the SCDi for all of Africa established that over the 1997-2024 reference period, political violence typically emerged in a low-intensity pattern, with roughly 67% of new conflict cells entering as clustered low-intensity (CL) or dispersed low-intensity (DL) (Radil et al., 2026). In 2025 for North and West Africa, the combined proportion of low-intensity entries was 56% (35% CL and 21% DL).

**Figure 6.** Initial states of newly affected conflict cells, 2025 in North and West Africa versus 1997-2024 historical average for all of Africa

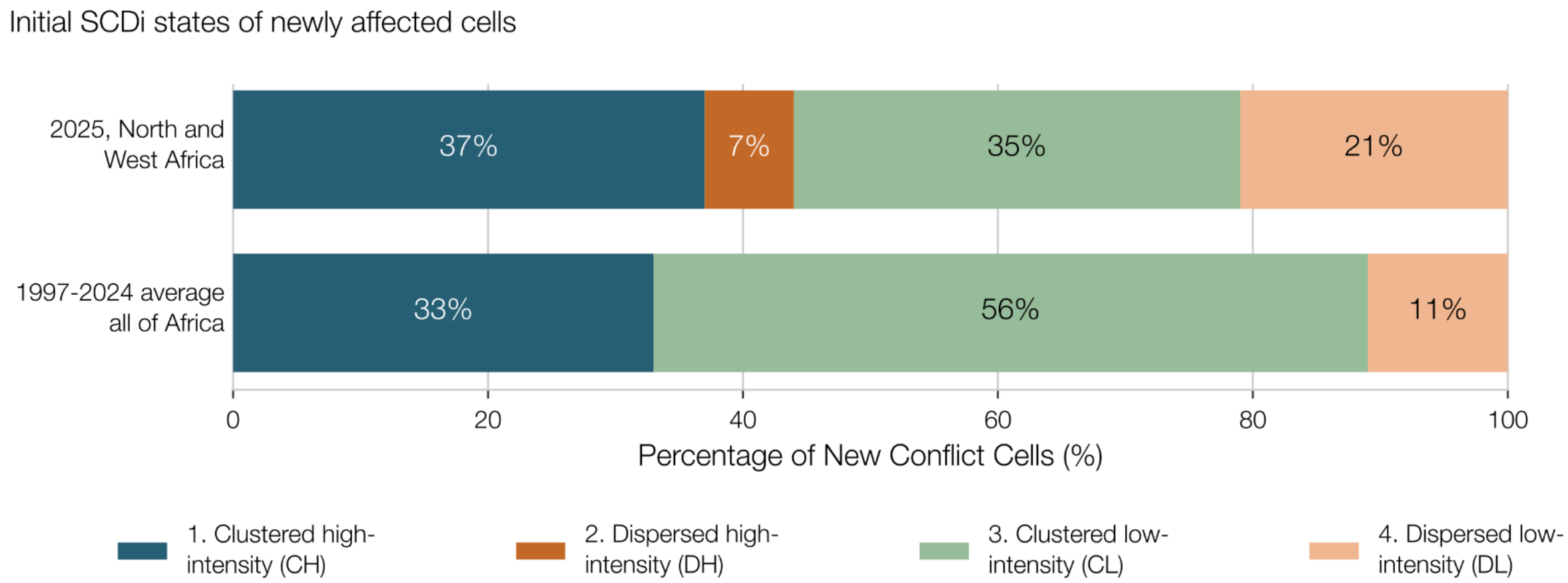

Source: Authors, based on ACLED (2026) and Radil et al. (2026).

Conversely, direct escalation into high-intensity violence has accelerated. Historically, approximately 33% of newly affected locations entered directly as clustered high-intensity (CH) conflict. In 2025, that figure rose to 37%, alongside a 7% entry rate for dispersed high-intensity (DH) conflict (Figure 6). This deviation indicates that peripheral areas are increasingly being absorbed into mature conflict zones at full force, rather than passing through a gradual, low-intensity transitional phase. The geography of expansion is shifting from diffuse, low-level spread into peripheral areas toward the rapid outward growth of dense, established conflict clusters.

Among cells that remained in conflict across both years, the flow between types was strongly upward on the intensity dimension. The most common transition was from CL to CH, representing the escalation pathway from peripheral to core conflict. The reverse movement, from CH to CL, occurred at roughly half the rate. Combined with the direct high-intensity entry of new cells, this pattern accounts for the continued growth of the CH category even as the absolute number of CL cells has remained stable.

## Identifying Local Conflict Trends

The local SCDi metrics, introduced recently, extend the base indicator by comparing current conditions in each cell against that cell's own history (Radil and Walther, 2024). Where the base SCDi classifies cells using region-wide thresholds, the local metrics ask whether violence in a specific location is currently more or less intense, and more or less concentrated, than it has been in the past. These metrics also identify cells that are newly entering conflict and track how many years of conflict each cell has accumulated.

*The geography of changing violence in 2025*

Map 3 presents the local SCDi results for 2025, combining two kinds of information in a single display. The filled circles show whether cells with established conflict histories are experiencing more or less intense, and more or less clustered, violence in 2025 when compared with their own 20-year average. The hollow circles mark cells that entered conflict in 2025 with no record of violence in the preceding 20 years.

**Map 3.** Local changes in conflict intensity and spatial clustering in North and West Africa, 2025

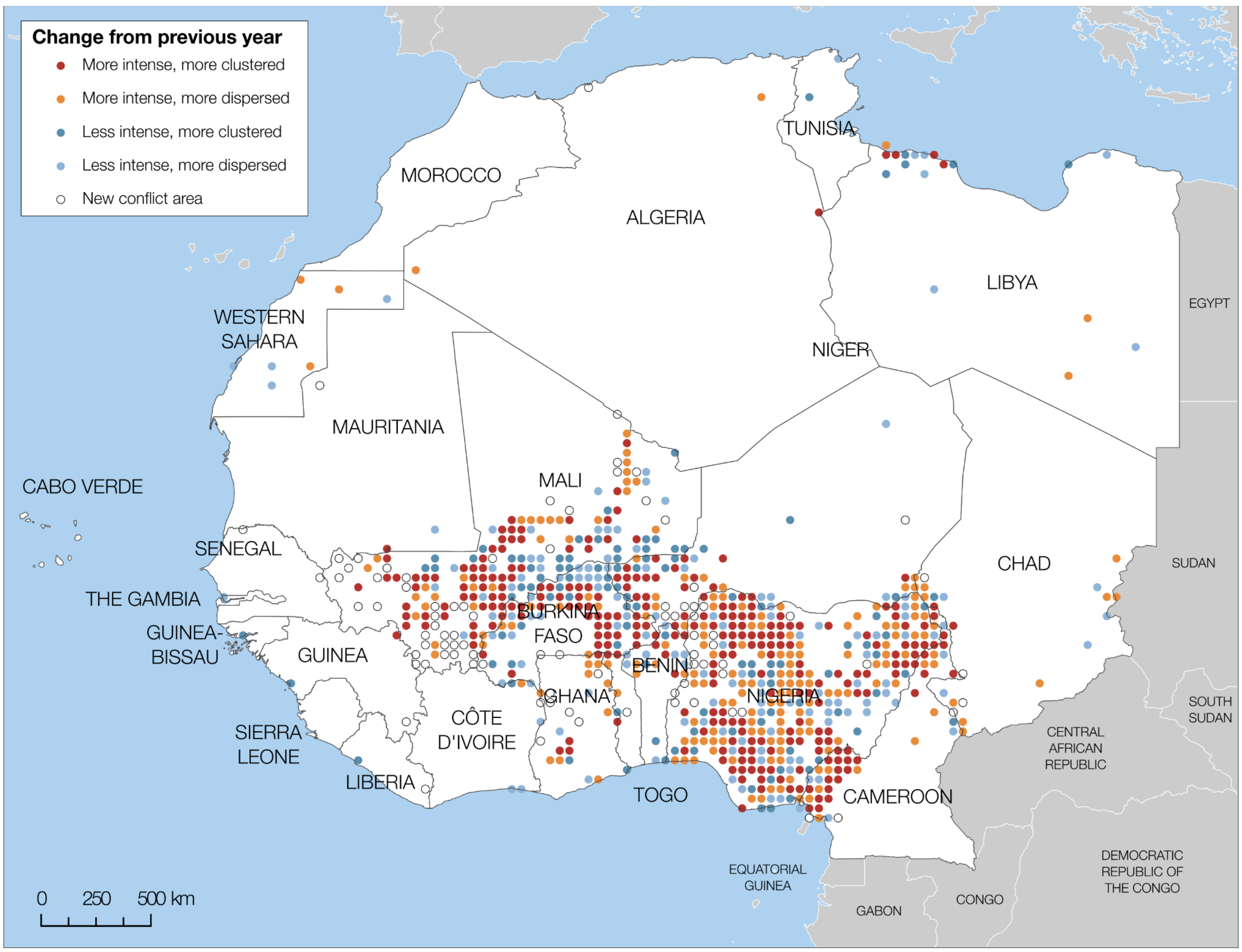

Source: Authors, based on ACLED (2026).

*The persistence of violence*

The geographic pattern is revealing. The core conflict zones of the Central Sahel and Nigeria are dominated by cells where violence in 2025 was both more intense and more clustered than in the past. These regions are visible in the dense concentrations of red across Burkina Faso, Mali, western Niger, and the Nigerian Middle Belt and Northwest. These are locations where conflict has been present for years and is continuing to deepen. At the periphery, the pattern shifts. Hollow circles marking cells with no prior 20-year history appear throughout southern Mali, within the states of northwestern Nigeria, and, to a lesser extent, across Ghana. The sources of conflict vary by location but all these newly affected areas are experiencing violence for the first time in a generation.

Among the 717 cells with sufficient prior conflict history to compute local scores, nearly two-thirds recorded higher intensity in 2025 than their historical baseline. In 2023, the equivalent proportion was 56% (Radil and Walther, 2024). The concentration metric presents a more stable picture: slightly more than half of cells were more clustered in 2025 than their historical average, a proportion essentially unchanged since 2023. The combination of rising intensity with stable concentration indicates that violence is becoming more frequent in familiar locations without fundamentally changing its within-cell spatial patterning.

*Persistence and conflict trajectories*

The years-in-conflict metric tracks how many of the 20 years in the reference window (2005-2024) each cell recorded qualifying violent events. A typical conflict cell in 2025 had experienced violence in about five of the past twenty years, but the distribution is concentrated at both extremes. At one end, 88 cells (11%) had no prior conflict history, representing entirely new episodes. At the other, 135 cells (17%) have now accumulated ten or more years of conflict, a group that has grown substantially from 82 cells (8%) in 2023 (Radil and Walther, 2024). Of the 29 cells recorded conflict in 15 or more of the past 20 years, all but two are in West Africa.

Twenty of these 29 cells are in Nigeria, spread across every major conflict zone in the country, from the Niger Delta through the northeast and northwest to the Middle Belt and the southeast. Mali follows with 3 cells, Libya with 2, and Chad, Côte d'Ivoire, Guinea, and Sierra Leone with 1 each. Nigeria's dominance at the upper extreme of the persistence distribution reflects two decades of overlapping conflicts with distinct origins and actors, a depth of layered violence that no other country in the region matches. Despite their prominence in current policy discussions, the Central Sahel contributes only three cells to this group (all in Mali: Kidal, Ménaka, Timbuktu), a reminder that the Sahelian crises remains a younger conflict system than Nigeria's and that its violence is typically ruralised. Most of these 29 cells also contain national or regional capitals, where political contestation, election-related violence, and security operations combine to produce a recurring pattern of violence that the persistence metric captures alongside the sustained insurgencies in the Sahel, even though the two differ in character.

A recent longitudinal analysis of conflict trajectories across Africa (Radil et al., 2026) provides context for interpreting this persistence. That study found that conflicts follow distinct 'life cycles,' ranging from ephemeral outbreaks that resolve within a year or two to enduring patterns where high-intensity clustered violence recurs for a decade or more with very low probabilities of de-escalation. The 135 cells with ten or more years of conflict in the current dataset correspond to these longer-term trajectory types. Their defining characteristic is a very high rate of self-reinforcement: once a cell enters a sustained clustered and high-intensity pattern, the probability that it will transition to a lower-intensity state or exit conflict entirely drops sharply. The growth of this group, from 8% to 17% of the conflict map in two years, reflects the consolidation of violence in the Central Sahel and Nigeria into patterns that have become self-sustaining.

The 88 cells at the opposite extreme face a different set of probabilities. The trajectory analysis suggests that most cells entering conflict for the first time follow ephemeral pathways, with violence appearing briefly and resolving within one or two years. The peripheral scatter of new conflict cells visible in Map 3 is consistent with this pattern. Whether any of these cells transition onto longer-term trajectories will depend on local conditions and on the proximity of established conflict zones, since the absorption of adjacent cells into existing clustered and high-intensity regions, documented earlier in this report, is one mechanism through which ephemeral outbreaks can become entrenched.

The local metrics describe a conflict map defined by two parallel dynamics. At the margins of established conflict zones, violence continues to reach new locations, most of which have no recent experience of political violence. In the core, locations are intensifying relative to their own historical baselines, and a growing share of the map is occupied by cells with a decade or more of accumulated conflict where the probability of de-escalation is low. The conflict landscape is widening at its edges and hardening at its centre.

## Outlook for 2026

The preceding sections of this report analysed the changing geography of violence through the end of 2025 using complete annual data. Because the SCDi requires a full calendar year of events to classify each cell, the indicator cannot be updated for 2026 until that year's data is complete. However, partial-year data from ACLED through May 2026 can be used to project full-year event and fatality totals for 2026 at the country level, providing an initial indication of whether the trajectories observed in 2025 are continuing.

Full-year projections are generated using a forecasting model fit to monthly country-level data from January 2020 through May 2026.[2] The model predicted monthly totals for the remaining seven months of 2026 (June through December), which were summed with the year-to-date actual data to produce projected annual totals. Events and fatalities were forecast independently, since per-event lethality varies by country and over time.

The projections suggest that the total number of violent events across the 21 study countries will reach approximately 15,400 in 2026. If realised, that would be a 17% increase over the roughly 13,200 events recorded in 2025. Similarly, projected fatalities of approximately 32,500 would represent a 20% increase over the 27,000 fatalities recorded in 2025. If these estimates hold, 2026 would mark a second consecutive year of acceleration following the relative plateau of 2023-2024 (Figure 7).

[2] The 2026 projections were generated by applying an Error, Trend, and Seasonality (ETS) exponential smoothing state space model (Hyndman and Khandaka, 2008) to historical monthly event and fatality data from January 2020 through May 2026. This time-series model automatically isolates underlying data noise, directional movements, and cyclical variations to accurately forecast the remaining seven months of the year for each country.

**Figure 7:** Violent events and fatalities in West Africa, 2020–2025 actuals and 2026 projections

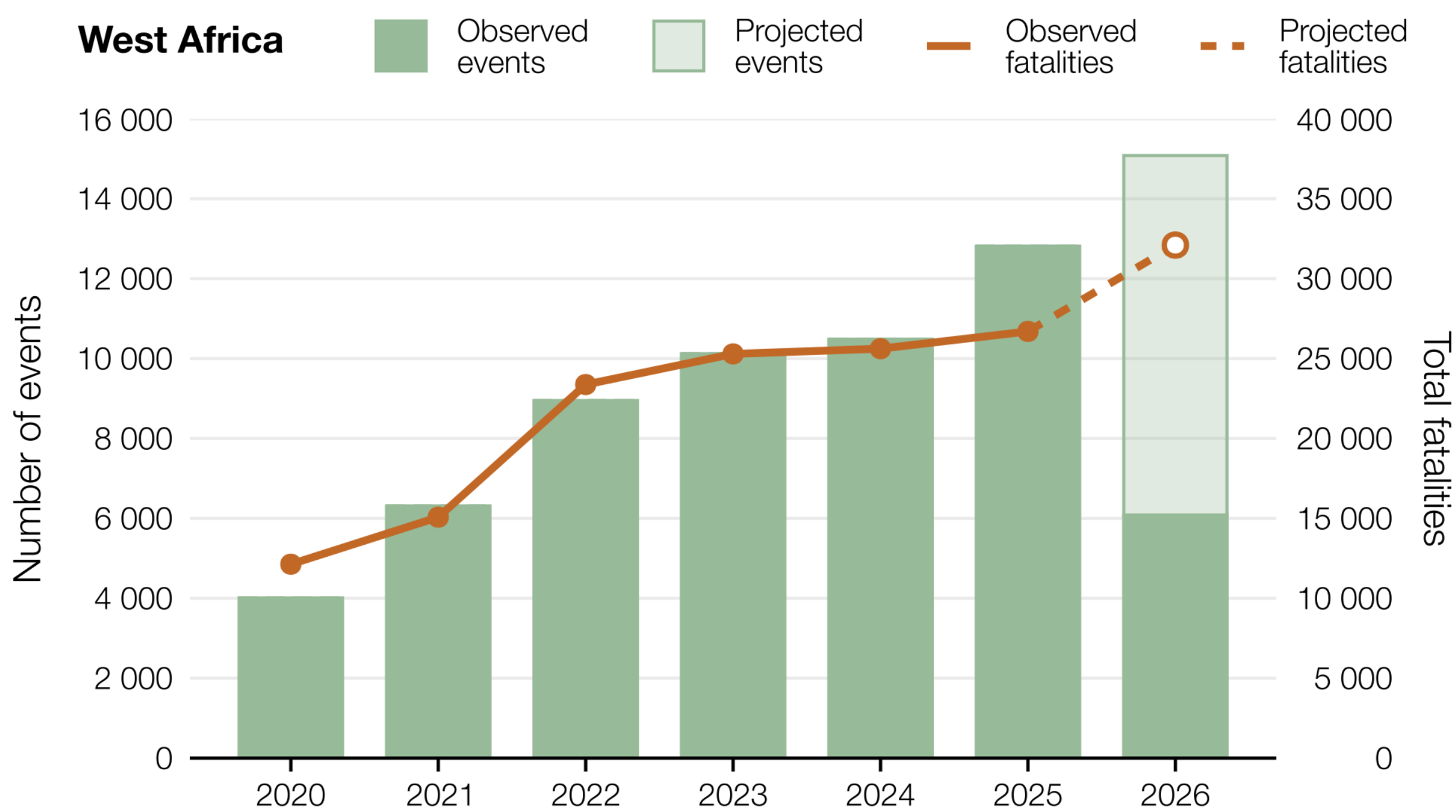

Source: Authors, based on ACLED (2026).
Note: Projected events and fatalities are estimated for the period from June through December, 2026 and combined with observed data from January through May.

The projected increase is driven overwhelmingly by Nigeria, which accounts for the majority of the regional growth in both events and fatalities (Table 2). Nigeria is also the only major country where per-event lethality is projected to increase rather than decline, with fatalities rising faster than events. Cameroon continues its upward trajectory in events, though its projected fatality increase is modest.

The Central Sahel presents a more complex picture than events alone would indicate. Mali and Burkina Faso are both projected to record essentially flat event totals, but their fatality trajectories have diverged sharply. Mali's projected fatalities would rise by more than 40% over 2025, producing a per-event lethality among the highest in the region. Given that the geographic extent of Mali's conflict is still expanding, the violence within that footprint may become significantly more deadly. Burkina Faso has moved in the opposite direction. Its projected fatalities would extend the 30% decline already recorded between 2024 and 2025, roughly halving the annual death toll in two years while incident counts remained constant.

**Table 2:** Violent events and fatalities across North and West Africa, 2025 and 2026 projections

| Country | Events 2025 | Events 2026 (projected) | Events change (%) | Fatalities 2025 | Fatalities 2026 (projected) | Fatalities change (%) |
|---|---|---|---|---|---|---|
| Nigeria | 5 244 | 7 106 | +35.5 | 12 630 | 17 943 | +42.1 |
| Cameroon | 3 401 | 3 978 | +17.0 | 1 857 | 1 971 | +6.1 |
| Mali | 1 584 | 1 578 | -0.4 | 3 928 | 5 585 | +42.2 |
| Burkina Faso | 1 362 | 1 377 | +1.1 | 5 263 | 3 834 | -27.2 |
| Niger | 530 | 561 | +5.8 | 1 905 | 2 039 | +7.0 |
| Libya | 352 | 273 | -22.4 | 276 | 344 | +24.6 |
| Ghana | 248 | 224 | -9.7 | 325 | 217 | -33.2 |
| Benin | 153 | 20 | -86.9 | 402 | 149 | -62.9 |
| Chad | 124 | 138 | +11.3 | 247 | 281 | +13.8 |
| Togo | 85 | 32 | -62.4 | 91 | 43 | -52.7 |
| Others | 141 | 114 | -19.1 | 135 | 104 | -23.0 |
| **Total** | **13 224** | **15 401** | **+16.5** | **27 059** | **32 510** | **+20.1** |

Source: Authors, based on ACLED (2026).
Note: Projected events and fatalities are estimated for the period from June through December, 2026 and combined with observed data from January through May.

The most notable shift concerns the coastal countries. Benin's projected totals would return to levels not seen since before 2020, a sharp reversal from the escalation documented earlier in this report. Togo shows a similar projected decline. Ghana's projected decrease is more moderate. If these figures hold through the end of the year, they would add further evidence that the southward move of jihadist is likely to be uneven. Whether the decline reflects effective government counter-measures, a strategic reorientation by armed groups, seasonal variation in the dataset, or some combination of these factors will require ongoing monitoring.

These projections carry important limitations. They are based on five months of 2026 ACLED data and assume that patterns observed in the early part of the year will persist through the remainder. Events and fatalities are forecast using independent models, which means the implied per-event lethality for some countries, particularly those with small event counts, may be unreliable. Country-level projections also cannot capture the sub-national shifts that the SCDi is designed to track: the question of where within countries violence is intensifying or spreading can only be answered when the full 2026 grid classification is produced.

With those caveats, the partial-year data points toward a 2026 in which Nigeria's conflicts drive regional totals to new highs, the Central Sahel diverges between a more lethal Mali and a less lethal Burkina Faso, and the coastal expansion slows or holds steady. The next SCDi update, using complete 2026 data, will provide the spatial resolution needed to assess whether these country-level shifts correspond to changes in the geographic structure of conflict or represent fluctuations within an otherwise expanding conflict map.

## Conclusion

The 2025 SCDi update describes a conflict landscape in North and West Africa that has grown both wider and deeper than at any previous point in the indicator's 29-year time series. The geographic extent of violence reached a new record, reversing a two-year stabilisation that previous assessments had tentatively identified as a plateau. At the same time, the character of violence within the conflict map has shifted. A larger share of affected cells than ever before consists of high-frequency, spatially concentrated conflict, while the proportion of cells experiencing transitional or sporadic violence has fallen to its lowest recorded level. The conflict map is expanding while also consolidating around a core of locations where violence has become self-reinforcing and where the probability of de-escalation, based on the historical trajectory evidence, is low.

The mechanism driving this consolidation has changed. In earlier years, conflict expanded primarily through the appearance of new low-intensity cells at the periphery, some of which escalated over subsequent years while others resolved. That pathway still operates, and most cells entering conflict for the first time continue to do so at low intensity. But a growing share of newly affected cells are now entering directly at high intensity, absorbed into adjacent conflict zones rather than undergoing gradual escalation. The practical consequence is that the boundary between conflict-affected and unaffected areas is becoming sharper. Locations that transition from no conflict classification to active conflict are increasingly doing so at high intensity, and the local metrics confirm that where violence exists, it is more intense relative to each cell's own historical baseline than at any point in the time series.

The southward expansion toward the Gulf of Guinea and the 2026 projections introduce a note of complexity into this otherwise unidirectional story. Several coastal countries show signs of declining violence in early 2026, while the core Sahelian conflicts have diverged, with Burkina Faso's fatality toll declining even as Mali's rises sharply. Nigeria's projected acceleration would drive regional totals to new highs. Whether these divergent trajectories signal the beginning of a structural shift or represent short-term fluctuations within an expanding conflict map will only become clear with additional data.

The SCDi's value lies in its ability to identify these dynamics at a spatial resolution unavailable from country-level statistics: where violence is intensifying, where it is appearing for the first time, and where it has persisted long enough to reshape the relationship between populations, territory, and the state. The 2025 data indicates that all three of these processes accelerated during the year. The full 2026 update will provide the resolution needed to assess whether the conflict landscape continues to widen and harden, or whether the partial reversals visible in some countries mark the beginning of a different trajectory.